\documentclass[usegraphicx,useAMS]{mn2e}
\begin{document}

\title[Cosmological Significance of LSB galaxies]{The cosmological significance
of Low Surface Brightness galaxies found in a deep blind neutral-hydrogen 
survey}

\author[R. F. 
Minchin et al.]{R.~F.~Minchin,$^1$\thanks{E-mail:Robert.Minchin@astro.cf.ac.uk}
M.~J.~Disney,$^1$ Q.~A.~Parker,$^2$ P.~J.~Boyce,$^3$ W.~J.~G.~de~Blok,$^1$
\newauthor G.~D.~Banks,$^1$\thanks{Now at BAE Systems} R. D. Ekers $^4$
K.~C.~Freeman,$^5$ D.~A.~Garcia,$^1$ B.~K.~Gibson, $^6$
\newauthor  M.~Grossi,$^1$ R.~F.~Haynes,$^7$ P.~M.~Knezek,$^8$ R.~H.~Lang,$^1$ 
D.~F.~Malin,$^9$
\newauthor R.~M.~Price,$^{10}$   M.~Putman,$^{11}$ I.~M.~Stewart,$^{12}$ 
A.~E.~Wright$^4$\\
$^1$ School of Physics and Astronomy, Cardiff University, 5 The Parade, 
Cardiff, CF24 3YB\\
$^2$ Department of Physics, Macquarie University, Sydney, NSW 2109, Australia\\
$^3$ Planning Division, Cardiff University, Park Place, Cardiff, CF10 3UA\\
$^4$ Australia Telescope National Facility, PO Box 76, Epping, NSW 1710, 
Australia\\
$^5$ Research School of Astronomy \& Astrophysics, Mount Stromlo Observatory, 
Cotter Road, Weston ACT 2611, Australia\\
$^6$ Centre for Astrophysics and Supercomputing, Swinburne University of 
Technology, PO Box 218, Hawthorn, Victoria 3122, Australia\\
$^7$ School of Mathematics \& Physics, University of Tasmania, Hobart, 
Tasmania 7001, Australia\\
$^8$ WIYN Consortium Inc., 950 North Cherry Avenue, Tucson, AZ 85719, United 
States\\
$^9$ Anglo-Australian Observatory, P.O. Box 296, Epping, NSW 1710, Australia\\
$^{10}$ Department of Physics and Astronomy, University of New Mexico, 
800 Yale Boulevard NE, Albuquerque, NM, United States\\
$^{11}$ Center for Astrophysics and Space Astronomy, University of Colorado, 
Campus Box 389, Boulder, CO, United States\\
$^{12}$ Department of Physics and Astronomy, University of Leicester, 
University Road, Leicester LE1 7RH}
\maketitle

\begin{abstract}
We have placed limits on the cosmological significance of gas-rich low 
surface-brightness (LSB) galaxies as a proportion of the total population
of gas-rich galaxies by carrying out a very deep survey (HIDEEP; Minchin et al.
2003) for neutral hydrogen (H{\sc i}) with the Parkes multibeam system.  Such a
survey avoids the surface-brightness selection effects that limit the 
usefulness of optical surveys for finding LSB galaxies.  To complement the
HIDEEP survey we have digitally stacked eight 1-hour $R$-band Tech Pan films 
from the UK Schmidt Telescope covering 36 square degrees of the survey area
to reach a very deep isophotal limit of 26.5 $R$\,mag\,arcsec$^{-2}$.  At this 
level, we find  that all of the 129 H{\sc i} sources within this area have 
optical counterparts and that 107 of them can be identified with individual 
galaxies.  We have used the properties of the galaxies identified as the 
optical counterparts of the H{\sc i} sources to estimate the significance of 
LSB galaxies (defined to be those at least 1.5 magnitudes dimmer in effective 
surface-brightness than the peak in the observed distribution seen in optical 
surveys).  Two different methods of correcting for ease-of-detection do not 
yield significantly different results: LSB galaxies make up $62 \pm 37$ per 
cent of gas-rich galaxies by number according to our first method (weighting 
by H{\sc i} mass function), which includes a correction for large scale 
structure, or $51 \pm 20$ per cent when calculated by our second method 
($1/V_{max}$ correction).  We also find that LSB galaxies provide $30 \pm 10$ 
per cent of the contribution of gas-rich galaxies to the neutral hydrogen 
density of the Universe, $7 \pm 3$ per cent of their contribution to the
luminosity density of the Universe; $9 \pm 4$ of their contribution to the 
baryonic mass density of the Universe, $20 \pm 10$ per cent of their 
contribution to the dynamical mass density of the Universe and $40 \pm 20$ per 
cent of their cross-sectional area.  We do not find any 
`Crouching Giant' LSB galaxies such as Malin 1, nor do we find a population of 
extremely low surface-brightness galaxies not previously found by optical 
surveys.  Such objects must either be rare, gas-poor 
or outside the survey detection limits.
\end{abstract}

\begin{keywords}
surveys -- galaxies: luminosity function -- radio lines: galaxies -- galaxies:
fundamental parameters
\end{keywords}
\section{introduction}
\label{intro-sec}
Surveys in the optical are known to be affected by strong selection effects
(e.g. Disney 1976; Disney \& Phillipps 1983; McGaugh 1996) which 
bias our understanding of the local galaxy population.  Corrections for these
selection effects can be made, but these are large and controversial and must
be applied to small numbers of sources, leading to large uncertainties (Impey
\& Bothun 1997; Disney 1999).  Low Surface-Brightness (LSB) galaxies could
make a significant contribution to the Universe that would not be recognised
in an optical survey.  They may dominate the luminosity, baryon or mass 
density of the galaxies in the Universe (e.g. Fukugita, Hogan, \& Peebles 1998)
and could contribute significantly to QSO absorption line spectra (e.g. 
Churchill \& Le Brun 1998).  

Zwaan et al. (2003) derived the H{\sc i} mass function for the 1000 
galaxies in the H{\sc i} Parkes All Sky Survey (HIPASS) Bright Galaxy 
Catalogue (Koribalski et al. 2004) and put a value of 15 
per cent on the contribution of LSB galaxies to the neutral hydrogen density of
the Universe.  However, this number depends crucially on their assumption that
optical catalogues are as (in)complete for LSB galaxies as for high 
surface-brightness (HSB) galaxies.  This is obviously not the case as LSB
galaxies are much harder to detect than HSB galaxies.  Indeed, most of the new 
galaxies, outside of the zone-of-avoidance, found in the HIPASS Bright Galaxy 
Catalogue are LSB galaxies (Ryan-Weber et al. 2002).  Thus the
value of Zwaan et al. should be considered only a lower limit to the 
contribution of LSB galaxies to the neutral hydrogen density.

The idea of searching for galaxies via the 21-cm neutral hydrogen line has long
been considered as an alternative to optical surveys (e.g. Disney 1976).  
However, until recent advances in technology such as multibeam receivers and 
powerful correlators, such a survey has not been practical.

The HIDEEP survey (Minchin et al. 2003; Paper 1 hereafter) was motivated 
by the desire to reach previously inaccessible
surface-brightness levels.  If optical surface-brightness (e.g. luminosity per
square arcsec) were to correlate with hydrogen column-density (e.g. H{\sc i}
flux per square arcsec), then to reach low surface-brightnesses it would be
necessary to reach low column-densities.  With an integration time of
9000 seconds per beam, HIDEEP is significantly more sensitive to low
column-density gas than any previous survey and thus potentially more sensitive
to low surface-brightness galaxies.

HIDEEP followed the same survey strategy and data-reduction path as HIPASS
(Barnes et al. 1998), but with twenty times the integration time.  The Parkes 
multibeam system was 
actively scanned in declination to cover a $6^\circ \times 10^\circ$ region, 
with full sensitivity being reached over the central $4^\circ \times 8^\circ$.
Scans were interleaved to give Nyquist sampling and repeated to reach the
required depth.  The data were treated in the same way as HIPASS data: they
were reduced in AIPS++ using the {\sc LiveData} and {\sc Gridzilla} programs 
and continuum sources were removed by template-fitting with the {\sc Luther}
program.  In order to minimise the noise, all the observations were carried
out at night.  The full sensitivity reached was 3.2 mJy beam$^{-1}$, which is
consistent with a $\sqrt{t}$ improvement on HIPASS.

To complement the deep H{\sc i} data, we obtained eight 1-hour exposures on
Tech Pan films at the UK Schmidt Telescope (UKST).  These covered a sky
area of $6^\circ \times 6^\circ$, with the same centre as the H{\sc i} survey
data.  The films were digitised using the SuperCOSMOS machine at the Royal 
Observatory,
Edinburgh (Hambly et al. 2001).  Digital stacking of these films brings a
$\sqrt{t}$ improvement in signal to noise (Bland-Hawthorn, Shopbell \&
Malin 1993), thus gaining us a little over a magnitude in depth.  The use of 
Tech Pan films gives a further magnitude improvement over the III-aF 
{\it R}-band 
plates previously used at the UKST (Parker \& Malin 1999).  For galaxies with 
ordinary colours, our $1\sigma$ surface-brightness limit of 26.5 {\it R} mag 
arcsec$^{-2}$ is equivalent to $\sim 27.5$ {\it B} mag arcsec$^{-2}$.

  The bandpass limit of the multibeam system
means that only objects with a heliocentric velocity less than 12,700 
km\,s$^{-1}$ can be detected (173 Mpc for H$_0$ = 75 km\,s$^{-1}$ 
Mpc$^{-1}$, after correction to CMB rest frame velocities), thus the volume 
of the 36 square degree optical overlap region is 19,000 Mpc$^3$.  Within this
volume we find a total of 129 sources, of which 107 can be identified with
individual galaxies.

\section{Analysis of the optical data}
\label{opt-sec}

In Paper 1, we identified sources found in the H{\sc i} data with optical
counterparts on the Tech Pan films.  A comparison of the offsets of the optical
positions from the H{\sc i} positions for the sources with secure 
identifications (those with catalogued velocities matching the H{\sc i}
velocity and those confirmed by H{\sc i} interferometry or optical
spectroscopy follow-up observations) with the offsets
for less certain counterparts (those without catalogued velocities or
follow-up observations) showed no significant difference.  Of the 107
galaxies in this paper, 87 are secure identifications; this includes 18 
confirmed with our interferometric follow-up and 2 with our spectroscopic 
follow-up.  Of the remaining 20, 18 are previously catalogued galaxies that
do not have redshifts in the literature and two are new detections.

\begin{figure}
\includegraphics[width=84mm]{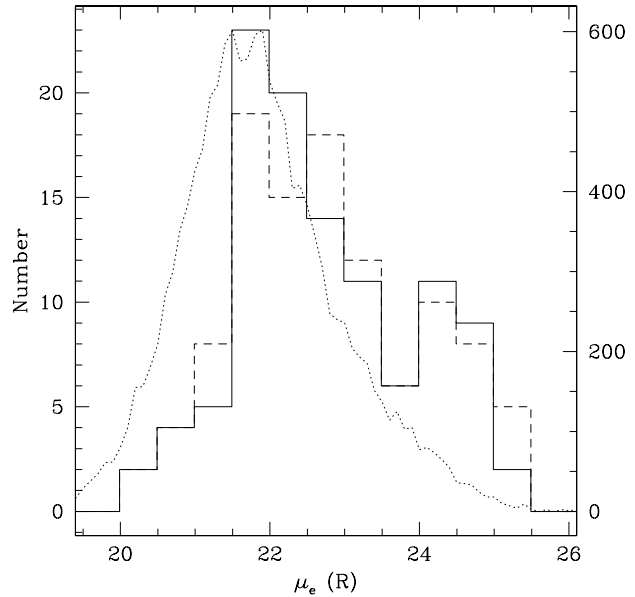}
\caption{Number of galaxies found in each surface-brightness bin.  The 
solid line shows the distribution of observed surface-brightnesses, the 
dashed line shows the distribution of surface-brightnesses after correction 
for galactic absorption, cosmological dimming, and inclination.  The dotted
line shows the observed surface-brightness distribution of ESO-LV galaxies 
(right hand scale).}
\label{sbhist}
\end{figure}

An H{\sc i}-selected sample is expected to appear different from an 
optically-selected one.  In particular low surface-brightness
galaxies -- which are seen over much smaller volumes than high 
surface-brightness galaxies with the same total luminosity in optical surveys 
(Disney \& Phillipps 1983) -- are expected to be seen much more easily.  
However, gas-poor galaxies,
such as ellipticals and dwarf spheroidals, which are found by optical surveys
will be invisible at 21-cm.  We find that this is indeed the case with
our sample.  As expected, our surface-brightness distribution (Figure 
\ref{sbhist}) is higher towards lower surface-brightnesses
than one finds in optically selected samples such as the ESO-LV (Lauberts
\& Valentijn 1989), which is also shown in the figure. A Kolmogorov-Smirnov 
test on 
our observed surface-brightness distribution and the surface-brightness 
distribution of the ESO-LV, which also uses {\it R}-band effective 
surface-brightness, shows 
that the hypothesis that both are drawn from the same parent population has 
a significance of less than 1 per cent, due to the larger
number of LSB galaxies seen in the HIDEEP sample.  This confirms that H{\sc i}
surveys do, as expected, avoid some of the surface-brightness selection
effects present in optical surveys.

\subsection{Tabulated Optical Properties}

\begin{table*}
\rotatebox{270}{\resizebox{126mm}{!}{\includegraphics{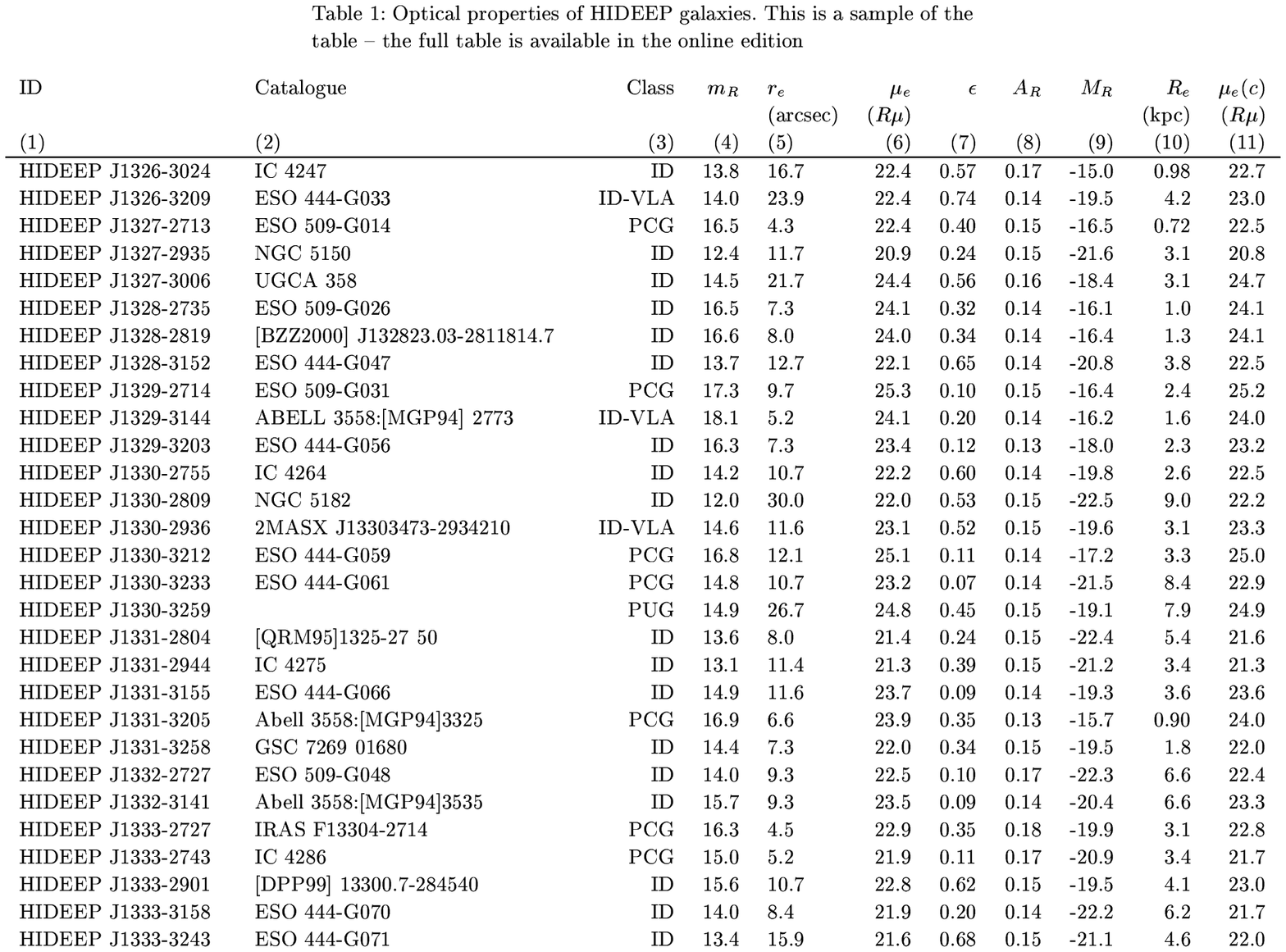}}}
\end{table*}
\addtocounter{table}{1}

Optical photometry of HIDEEP galaxies was carried out using a combination of 
{\sc sextractor} (Bertins \& Arnouts 1996) and the {\sc stsdas} package in 
{\sc iraf}, where the {\sc ellipse} task was used to obtain radial 
surface-brightness profiles.  The measured optical parameters are displayed 
in Table {1} (a sample is shown here, the full table is available in the 
online edition) where column (1) gives the source name and (2) the 
catalogued identification -- if there is one. Column 
(3) gives the type of the identification graded into 3 classes: 
``ID'' denotes objects where both positions and velocities coincide, ``PCG'' 
(for Previously Catalogued Galaxy) where there is a positional coincidence 
with a previously catalogued galaxy in the NASA Extragalactic Database (NED) 
and ``PUG'' (for Previously Uncatalogued Galaxy) where there is no possible 
counterpart in NED, and thus no previous positional or redshift data.  Three 
further sub-classes,  ``ID-VLA'', ``ID-ATCA'', and 
``ID-Spec'' denote where a previously catalogued galaxy without a redshift
has been confirmed as the optical counterpart through our follow-up 
observations (by interferometry with the VLA or the ATCA or by spectroscopy at 
the ANU 2.3-m telescope at Siding Spring respectively).

The tabulated optical parameters are as follows: column (4) gives the apparent
magnitude ($m_R$) as measured by the {\it mag}\_{\it best} parameter in {\sc 
sextractor}.  This takes the form of either an adaptive aperture magnitude
(Kron 1980), where the size of the aperture is 2.5 times the first-moment
radius of the galaxy, or, for crowded fields, an isophotal magnitude (to the 
26 $R$ mag arcsec$^{-2}$ isophote) with a correction 
made for the part of the galaxy beyond the isophotal limit (Maddox et al. 
1990), the best algorithm being selected automatically by the program.  
Simulations by Bertin \& Arnouts (1996) of an {\it R}-band CCD image
with an isophotal limit of 26 {\it R} mag arcsec$^{-2}$ (i.e. very similar
to our image)  show that, to the magnitude level of the faintest 
galaxy in our survey (18.6 {\it R} mag), the true magnitude is recovered to 
better than 0.1 mag. 

Column (5) gives the effective radius, $r_{e}$ -- the 
radius enclosing half the {\it R}-band light.  Column (6) gives the effective 
surface-brightness $\mu_{e}$ -- the {\it R}-band surface-brightness at the 
effective radius. This gives a 
model-independent measure of the surface brightness which can be applied to 
all types of galaxies.  Column (7) gives the {\sc sextractor} ellipticity 
($\epsilon = 1-b/a$) and (8) the estimated {\it R}-band absorption at the 
position of the object from Schlegel et al. (1998) supplied by NED.  Column
(9) gives the absolute magnitude $M_R$, including corrections for galactic 
absorption, cosmological dimming and internal absorption (estimated using 
$A_{i,R} = 0.95 \log (a/b)$, where $a/b$ is calculated from $\epsilon$).  
The k-correction out to the bandpass limit of 12,700 km\,s$^{-1}$ is 
negligible. We use the distances from Paper 1 except for galaxies in the
Centaurus A group (taken to be those within 1000 km\,s$^{-1}$), where we use 
the distance to the M83 subgroup of 4.5 Mpc from Thim et al. (2003).  Column 
(10) gives the physical effective radius $R_{e}$ in kpc.  Finally column (13) 
gives the surface brightness $\mu_{e}(c)$ corrected for $(1+z)^4$ cosmological 
dimming, inclination (estimated using $C_{i,R} = -1.25 \log (a/b)$ from Graham 
\& de Blok (2001)) and galactic absorption.

Table \ref{optpos} gives the positions of the optical counterparts of the PUGs.
Where these have been confirmed by interferometric follow up, this is indicated
by a code in column (4) giving the run in which they were detected.  The PUGs
that were not detected in earlier runs were followed up again in later runs, 
with the result that there are no galaxies in the table that have been 
followed-up but not detected, although there are two sources which have not
been followed-up interferometrically.

\begin{table}
\caption{Positions of newly-discovered galaxies in the HIDEEP optical
area.  For those detected interferometrically, Column (4) gives the code for 
the observing run: A1 - ATCA, 11/1999; A2 - ATCA, 01/2000; V - VLA, 01-03/2003;
A3 - ATCA, 04/2003.  Those without a code have not been observed with 
an interferometer.}
\label{optpos}
\begin{tabular}{llll}
HIDEEP ID&R.A.&Decl.&Run\\
(1)&(2)&(3)&(4)\\
\hline
J1330-3259&13:30:32&-32:57:07&V\\
J1336-2932&13:36:08&-29:34:12&A1\\
J1337-3118&13:36:59&-31:19:05&V\\
J1338-3035&13:37:56&-30:35:12&A2\\
J1339-3022&13:39:08&-30:22:10&V\\
J1342-2859&13:42:05&-29:01:21&A2\\
J1342-3021&13:42:21&-30:23:14&V\\
J1344-3202&13:44:03&-32:02:11&-\\
J1345-2908&13:45:30&-29:06:47&V\\
J1345-3041&13:45:20&-30:43:21&-\\
J1345-3104&13:45:12&-30:56:52&V\\
J1347-2735&13:47:37&-27:35:22&V\\
J1347-2810&13:47:45&-28:11:56&A3\\
J1348-2856&13:48:38&-29:00:22&V\\
\end{tabular}
\end{table}

\section{Volumetric and Large Scale Structure corrected data}
\label{volume-sec}

\begin{figure}
\includegraphics[width=84mm]{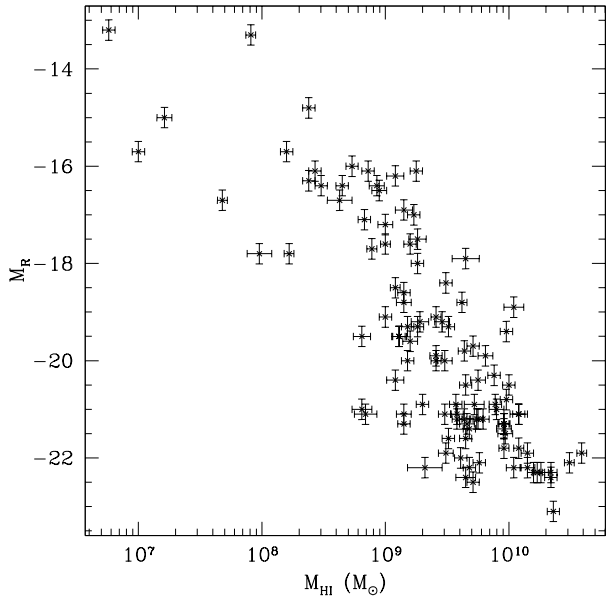}
\caption{Relationship between H{\sc i} mass and absolute magnitude}
\label{mhi-absmag}
\end{figure}

\begin{figure}
\includegraphics[width=84mm]{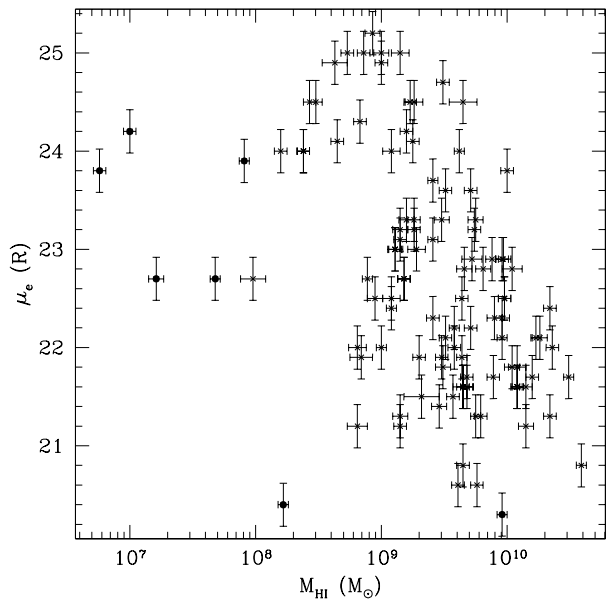}
\caption{Relationship between H{\sc i} mass and effective surface-brightness.  
Cen A group galaxies are shown by solid circles.  The relationship is weaker
than for luminosity, but there is a definite trend for lower H{\sc i} mass 
galaxies to have lower surface-brightnesses: 14 of the 23 galaxies with $M_{HI}
< 10^9 M_\odot$ are LSB galaxies ($\mu_e (R) > 23.3$ {\it R} mag 
arcsec$^{-2}$), while, at the high-mass end, there is only one LSB galaxy out
of the sixteen with $M_{HI} > 10^{10} M_\odot$.  This dependence of 
surface-brightness on H{\sc i} mass means that corrections must be made
for H{\sc i} selection effects before the surface-brightness distribution
can be determined.}
\label{mhi-sb}
\end{figure}

No conclusions can be drawn as to the cosmic significance of the LSB galaxies 
found in  HIDEEP until the raw numbers have been corrected for 
ease-of-H{\sc i}-detection and for the large scale structure in the region.  
The peak
flux ($S_{peak}$) of a galaxy determines the distance to which it may be seen 
in an H{\sc i} survey and is related to its H{\sc i} mass ($M_{HI}$) and its 
velocity width ($\Delta V$) as $M_{HI} \propto d^2 \Delta V S_{peak}$.  
However, the velocity width is not independent of the mass of a galaxy -- 
these are related as $\Delta V \propto M_{HI}^{1/3}$ (Briggs \& Rao 1993), 
thus $M_{HI} \propto d^3 S_{peak}^{3/2}$.  For a fixed limiting peak flux,
we therefore get $M_{HI,lim} \propto d^3$.  While this is an approximate
empirical relationship originally defined using an optically-selected sample
of galaxies, the general result that higher H{\sc i}-mass galaxies can be seen
to greater distances is seen to hold true in blind H{\sc i} surveys (e.g.
Zwaan et al. 1997 or Koribalski et al. 2004).  The exact form of this
dependence of limiting H{\sc i} mass on distance is not used in our analysis 
and is thus unimportant for the purposes of this study.

\begin{figure}
\includegraphics[width=84mm]{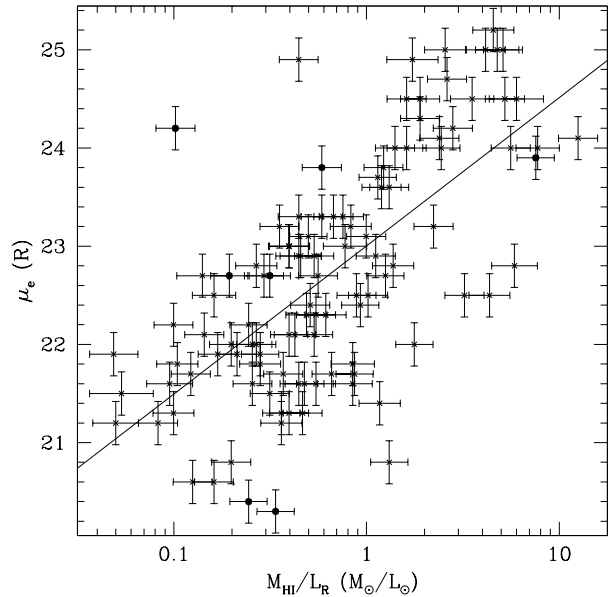}
\caption{Correlation of effective surface-brightness with H{\sc i} mass to 
light ratio.  Cen A group galaxies are shown by solid circles.  The correlation
between $\mu_e$ and $\log(M_{HI}/L_R)$, shown as a solid line, has a slope of 
$1.51 \pm 0.16$ and a scatter of 0.9 magnitudes.  As the survey detects 
galaxies by their H{\sc i} content, galaxies with low H{\sc i} mass-to-light
ratios may be missed, however this survey goes deep enough to see galaxies
down to $M_{HI}/L_R \simeq 0.05 M_\odot/L_\odot$ so it is unlikely that this
will be a significant effect.}
\label{mol-sb}
\end{figure}

We find that parameters such as absolute magnitude and surface brightness are 
correlated with hydrogen mass (Figures \ref{mhi-absmag} and \ref{mhi-sb}).  
Thus more luminous galaxies 
may be seen over larger volumes and, in a reversal of the situation in the
optical, low surface-brightness galaxies may be seen over larger volumes than
higher surface-brightness galaxies of the same luminosity due to the 
anti-correlation between surface-brightness and the H{\sc i}-mass to light 
ratio (Figure \ref{mol-sb}) seen both in our data and in previous studies 
using optically-selected samples (e.g.de Blok, McGaugh \& van der Hulst, 1996).

In order to make the volumetric corrections for H{\sc i} selection effects, we 
use two different methods,
which are described below.  We have chosen to remove dwarf galaxies ($M_{HI} 
< 10^8 M_\odot$) from this analysis for two main reasons: that the number of 
low-mass galaxies in our sample is low and that the volume in which we can
see these galaxies in dominated by the Cen A group.  For these reasons,
we focus solely on the more H{\sc i}-massive galaxies ($M_{HI} > 10^8 M_\odot$)
where we have reasonable statistics and probe a variety of environments.

\subsection{$1/V_{max}$ Weighting}

Weighting detections by $1/V_{max}$, where $V_{max}$ is the total volume in
which they could have been found, is well established as a means of correcting
for ease-of-detection.  We use this here with a completeness limit of 
18 mJy in peak flux (Paper 1) and a maximum distance, due to the bandpass limit
of the H{\sc i} survey, of 173 Mpc.  This gives us a complete sample of 67 
galaxies with $\left<V/V_{max}\right> = 0.49 \pm 0.04$.
Once the dwarfs are removed, we are left with 62 galaxies with 
$\left<V/V_{max}\right> = 0.52 \pm 0.04$.  While $1/V_{max}$ is well 
understood, it cannot make any correction for the large scale structure, which 
could lead to distortions in the results.  However the value of 
$\left<V/V_{max}\right>$ implies that the overall effect of large scale 
structure on the sample is probably not large.

\subsection{H{\sc i} Mass Function Weighting}

Our second method is to use an H{\sc i} Mass Function (HIMF) to correct for 
both ease-of-detection and large scale structure, as described in Minchin
(1999).  This method makes the assumption that our galaxies are drawn from the
same population as a general HIMF, such as that found for the HIPASS Bright
Galaxy Catalogue (BGC) by Zwaan et al. (2003).  Then the volumetric correction 
that needs to be applied in order to match the distribution of H{\sc i} masses 
in HIDEEP with the HIMF can be calculated and applied to find other quantities,
such as the luminosity function and the surface-brightness distribution.  
After dwarf galaxies are excluded, this gives us a sample of 101 galaxies.

This HIMF-weighting cannot be used to construct an H{\sc i} mass function for
the HIDEEP galaxies, as it would clearly just give the same answer as the
input HIMF.  However, we can use it to construct the bivariate brightness
distribution, luminosity function and surface-brightness distribution of
galaxies and to investigate how various parameters (e.g. luminosity density)
change with surface-brightness.  It would be possible to make a HIMF from 
the HIDEEP data using $1/V_{max}$, however we have chosen to use the HIMF
of Zwaan et al. (2003), based on the 1000 galaxies in the HIPASS BGC.  This 
contains an order of magnitude more galaxies than HIDEEP and thus gives a much 
more accurate mass function.

HIMF-weighting contains more sources of error than $1/V_{max}$.  The main
sources of additional errors are the width of the bins used in the HIMF and
the numbers of galaxies in each bin.  The resulting formal uncertainties are
therefore normally larger than for $1/V_{max}$, despite the larger sample
size.  However, it should be remembered that HIMF-weighting includes
a correction for the large scale structure which is not factored into either
the values or the errors of the $1/V_{max}$ weighting.  

\begin{table}
\caption{H{\sc i} Mass Functions used in calculating the weighting (corrected
to $H_0 = 75$ km\,s$^{-1}$\,Mpc$^{-1}$).}
\label{himfs}
\begin{tabular}{lllll}
$\alpha$&$\phi^\star$&log $M_{HI}^\star$&Reference\\
-1.30&0.0086&9.79&Zwaan et al. 2003 (BGC)\\
-1.53&0.005&9.88&Rosenberg \& Schneider 2002\\
-1.52&0.0032&10.1&Kilborn 2001\\
-1.51&0.006&9.7&Henning et al. 2000\\
-1.2&0.006&9.8&Zwaan et al. 1997\\
\end{tabular}
\end{table}

\begin{figure}
\includegraphics[width=84mm]{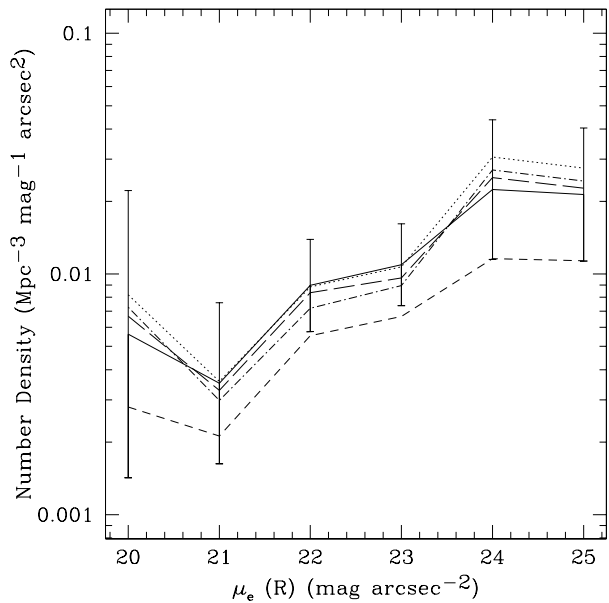}
\caption{The surface-brightness distribution formed by correcting with
various different H{\sc i} Mass Functions.  Solid line: HIPASS Bright Galaxy
Catalogue (BGC; Zwaan et al. 2003) as 
used elsewhere in this paper; dotted line: Rosenberg \& Schneider (2002);
short-dashed line: Zwaan et al. (1997); long-dashed line: Kilborn (2001); 
dot-dashed line: Henning et al. (2000).  Error bars are given for the
distribution formed using the BGC HIMF.}
\label{multihimf}
\end{figure}

Systematic errors may be introduced by the choice of the HIMF.  In order to 
gauge the size of these errors, we have looked at the contribution of
LSB galaxies to the total number density of gas-rich galaxies when calculated
with the HIMFs of Rosenberg \& Schneider (2002), Kilborn (2001), Henning et al.
(2000) and Zwaan et  al. (1997) (see Table \ref{himfs} for a summary).  The
resulting surface-brightness distributions are shown in Figure \ref{multihimf}.

It can be seen from Figure \ref{multihimf} that there is little 
difference in the overall shape of the Surface Brightness Distributions derived
using the different HIMFs, although the overall density of gas-rich galaxies is
significantly reduced if the HIMF of Zwaan et al. (1997) is used.
The contribution of LSB galaxies as a proportion of all galaxies is therefore 
only weakly dependant on the HIMF chosen, with a slight rise in the proportion
for steeper slopes.  Within the range of HIMFs here the proportion of LSB
galaxies varies between 59 and 67 per cent of all gas-rich galaxies, with 
the BGC HIMF giving 62 per cent.  As the statistical error on this proportion
is 37 per cent, it can be seen that the systematic error due to the
HIMF is comparatively
small.  These systematic errors are stated in Table \ref{importantLSB2} as
a second error column to the percentage LSB contribution.

Masters, Haynes, \& Giovanelli (2004) have shown that if a peculiar velocity
field model is used to assign distances to galaxies in the BGC rather than the 
pure Hubble law used by Zwaan et al. (2003) then the faint-end slope of the
BGC HIMF steepens from $\alpha = -1.3$ to $-1.4$, which is consistent with the 
steeper
determinations.  The HIMF of Zwaan et al. (1997) is also inconsistent with
the other determinations; this cannot be explained by the method used for
assigning distances but may be due to problems with correctly determining
the survey sensitivity (Schneider, Spitzak \& Rosenberg 1998).  It is most 
likely, therefore, that if a systematic error has been introduced by the use
of the HIMF, it is that the faint-end slope we have used is too shallow.  
The result of this would be for this paper to slightly underestimate the 
contribution of LSB galaxies.

\subsection{The Bivariate Brightness Distribution}

\begin{figure}
\includegraphics[width=84mm]{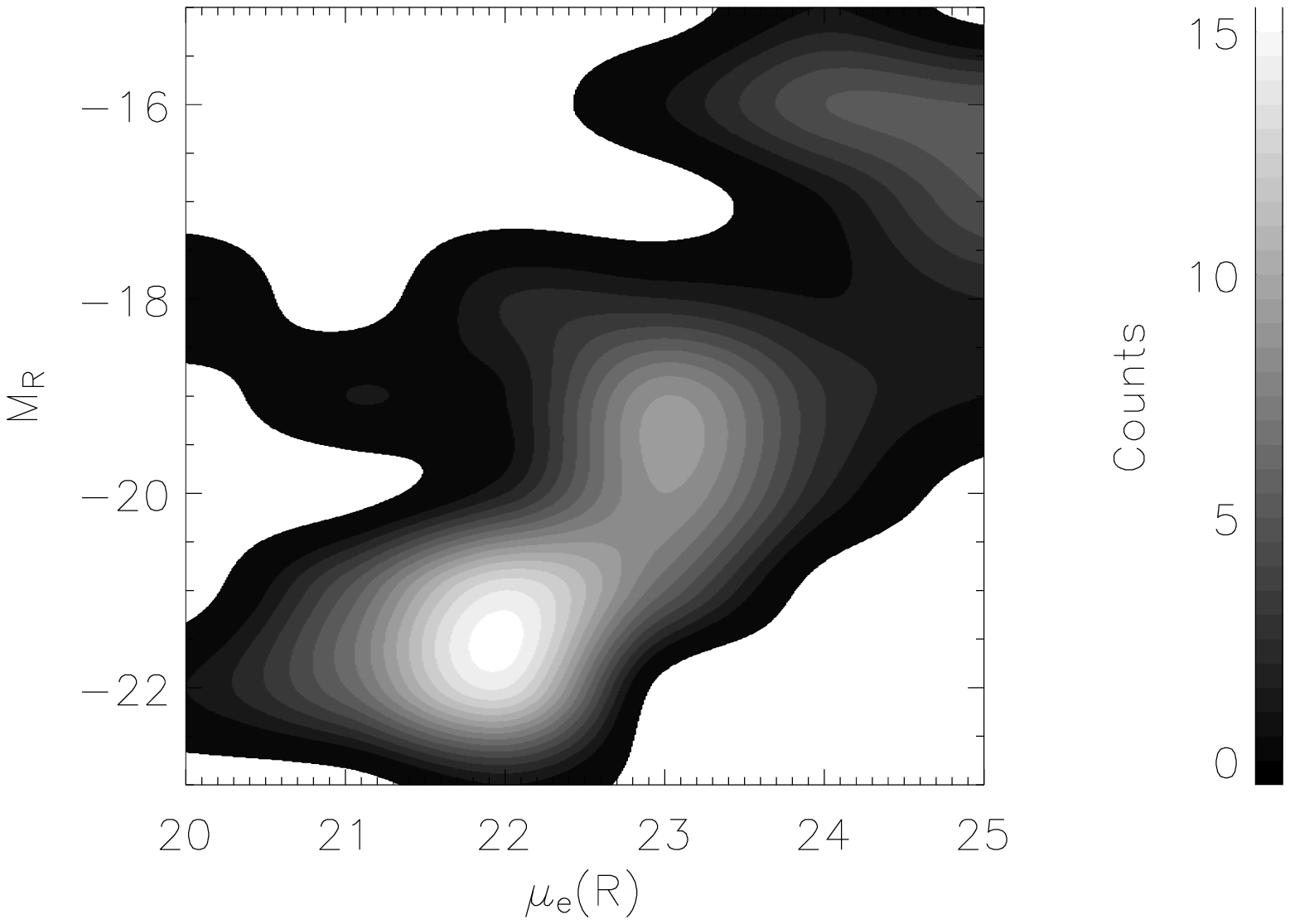}
\includegraphics[width=84mm]{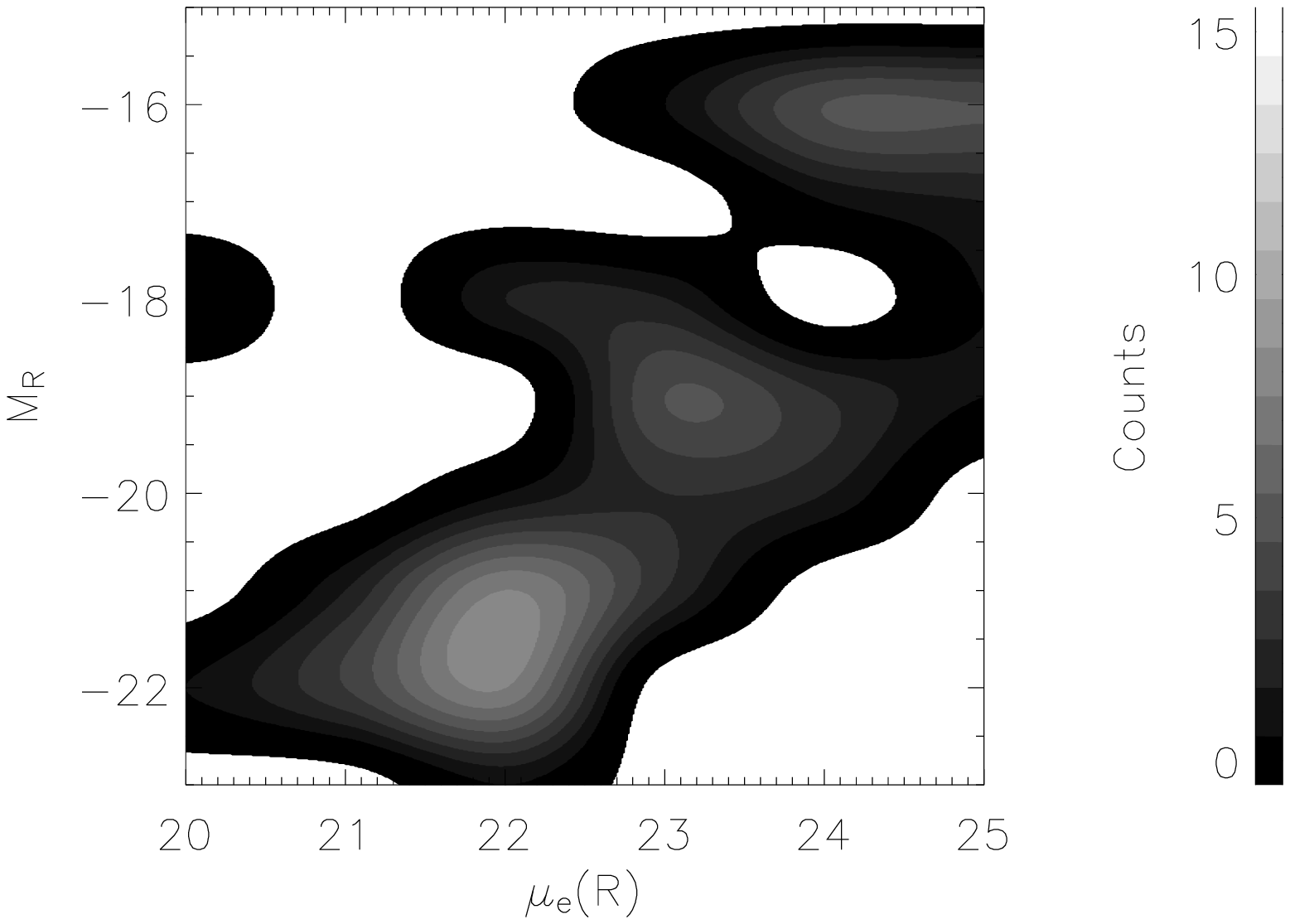}
\caption{Uncorrected bivariate brightness distribution of HIDEEP
galaxies.  The upper panel shows the 101 galaxies in the HIMF-weighted sample,
the lower panel the 62 galaxies in the $1/V_{max}$-weighted sample.  The blank 
areas around the images indicate where there is no data.}
\label{unwtbbd}
\end{figure}

As luminosity and surface-brightness appear to be correlated to some degree,
the distributions of 
luminosity and surface-brightness on their own are less interesting than the 
bivariate brightness distribution (BBD) -- the joint distribution in the 
($M_R$,$\mu_{e}^R$) plane.  The BBD describes the contribution of galaxies of 
different luminosities and surface-brightnesses to the cosmos but is virtually
impossible to obtain accurately from an optically-selected sample (Boyce \& 
Phillipps 1995).  For instance the largest sample of galaxies, complete in 
both surface brightness and luminosity, that can be assembled from the 
classical optical catalogues numbers less than 50 (Disney 1999).

Figure \ref{unwtbbd} shows the BBDs drawn from the uncorrected samples, showing
the correlation between luminosity and surface-brightness and
illustrating the conventional emphasis on the overwhelming importance of 
high-surface-brightness giant galaxies.  However these galaxies tend to 
have much larger H{\sc i} masses and can be seen over much greater volumes.

\begin{figure*}
\begin{minipage}{126mm}
\centerline{\includegraphics[width=84mm]{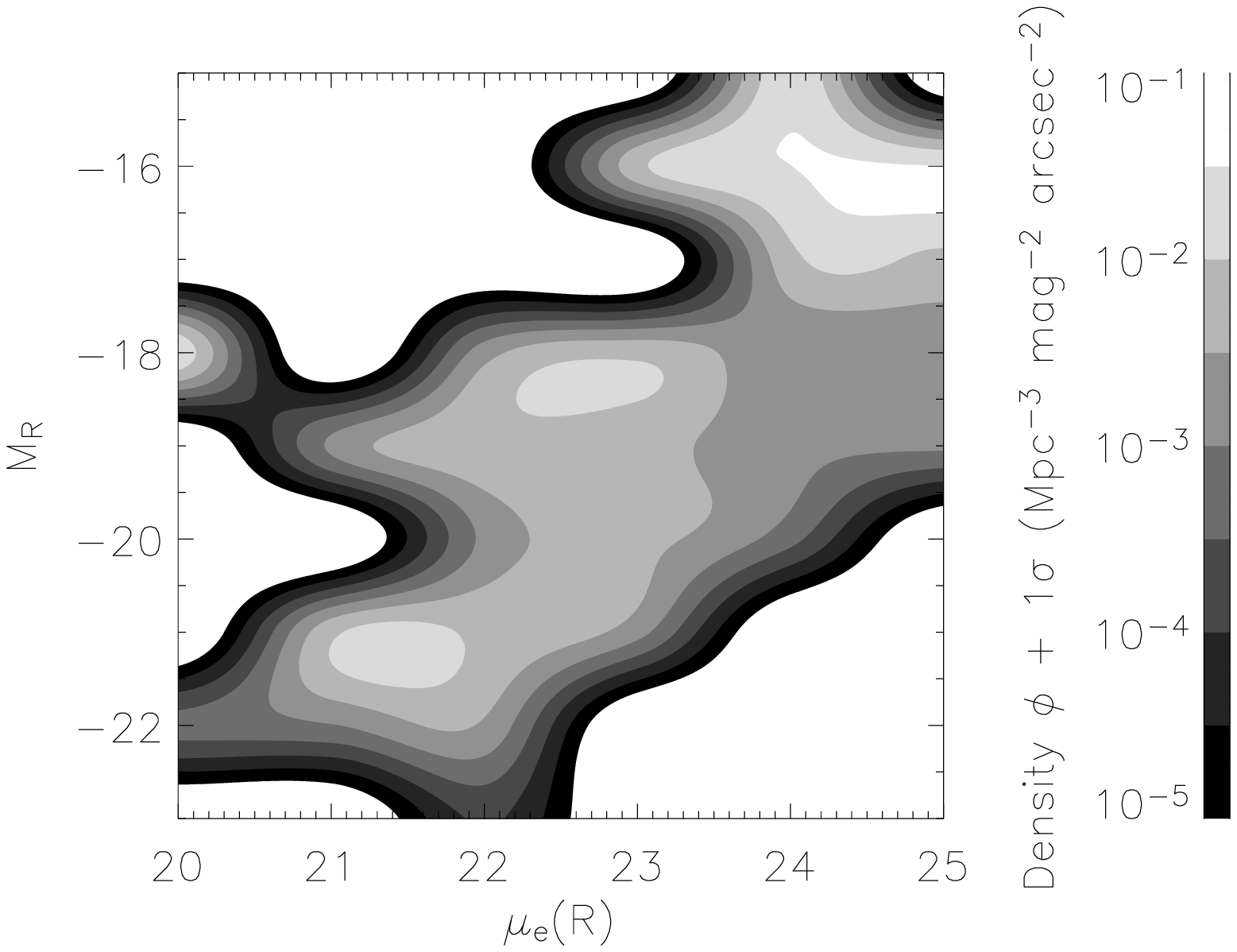}
\hfill
\includegraphics[width=84mm]{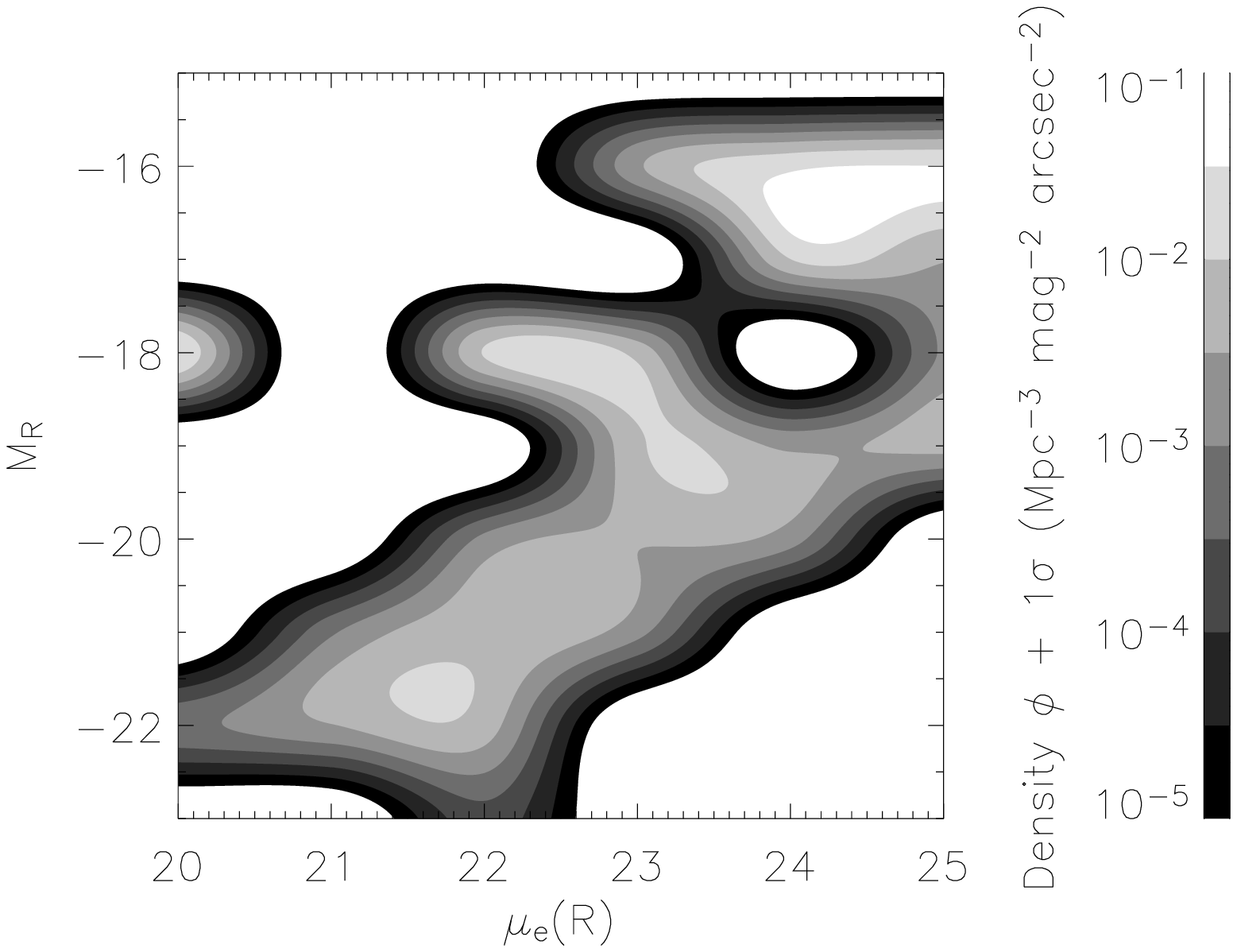}}
 
\centerline{\includegraphics[width=84mm]{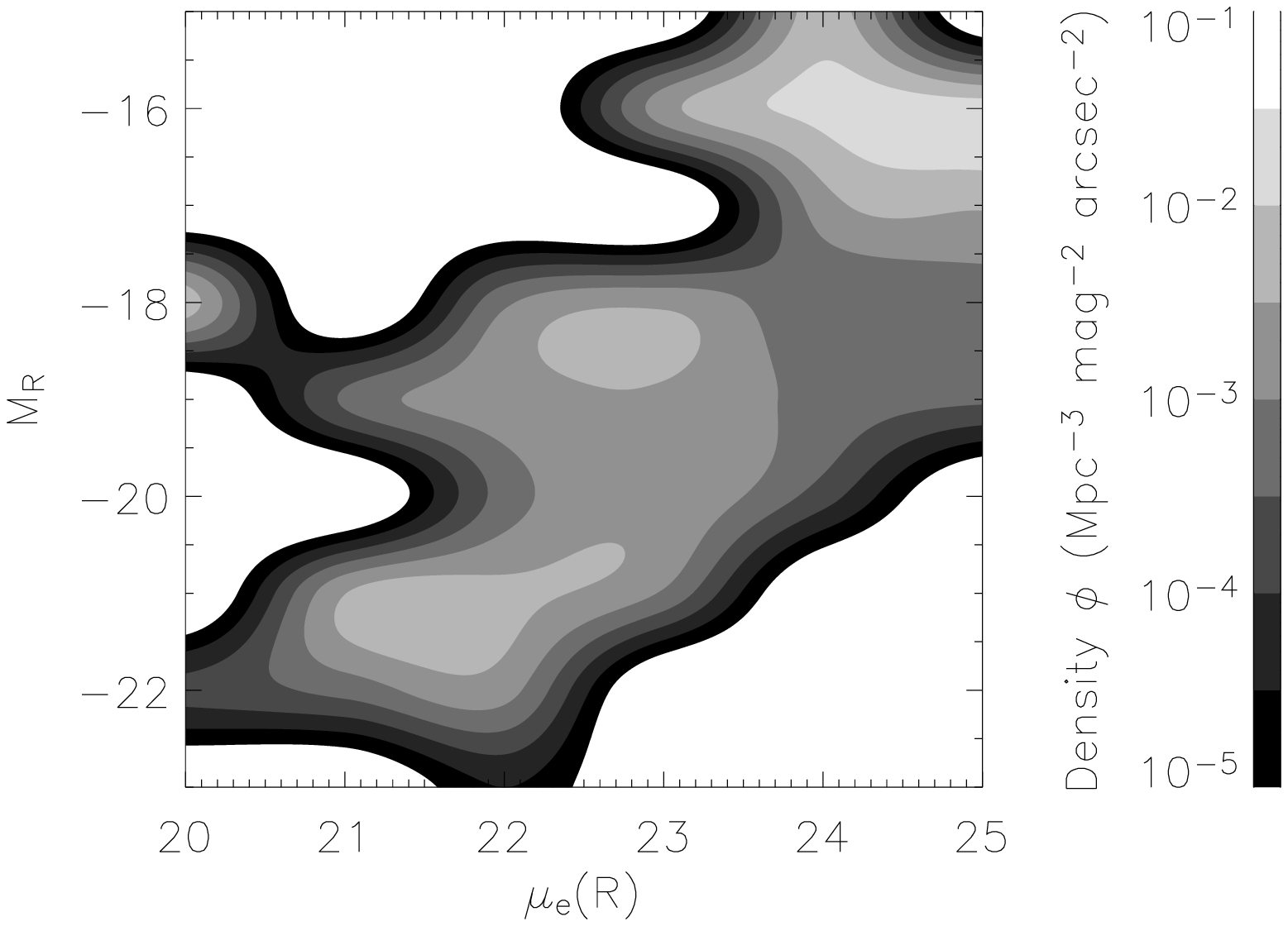}
\hfill
\includegraphics[width=84mm]{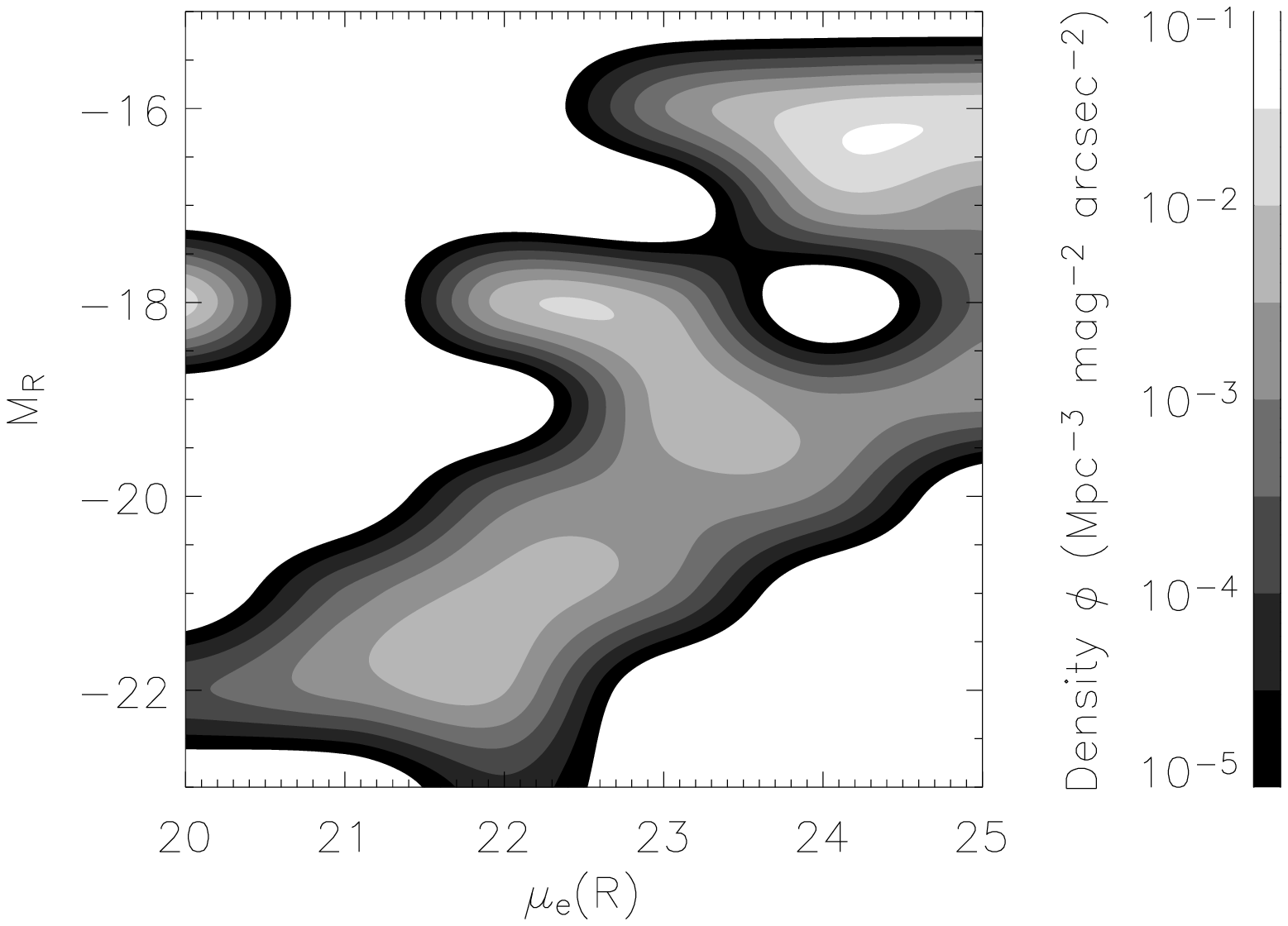}}
 
\centerline{\includegraphics[width=84mm]{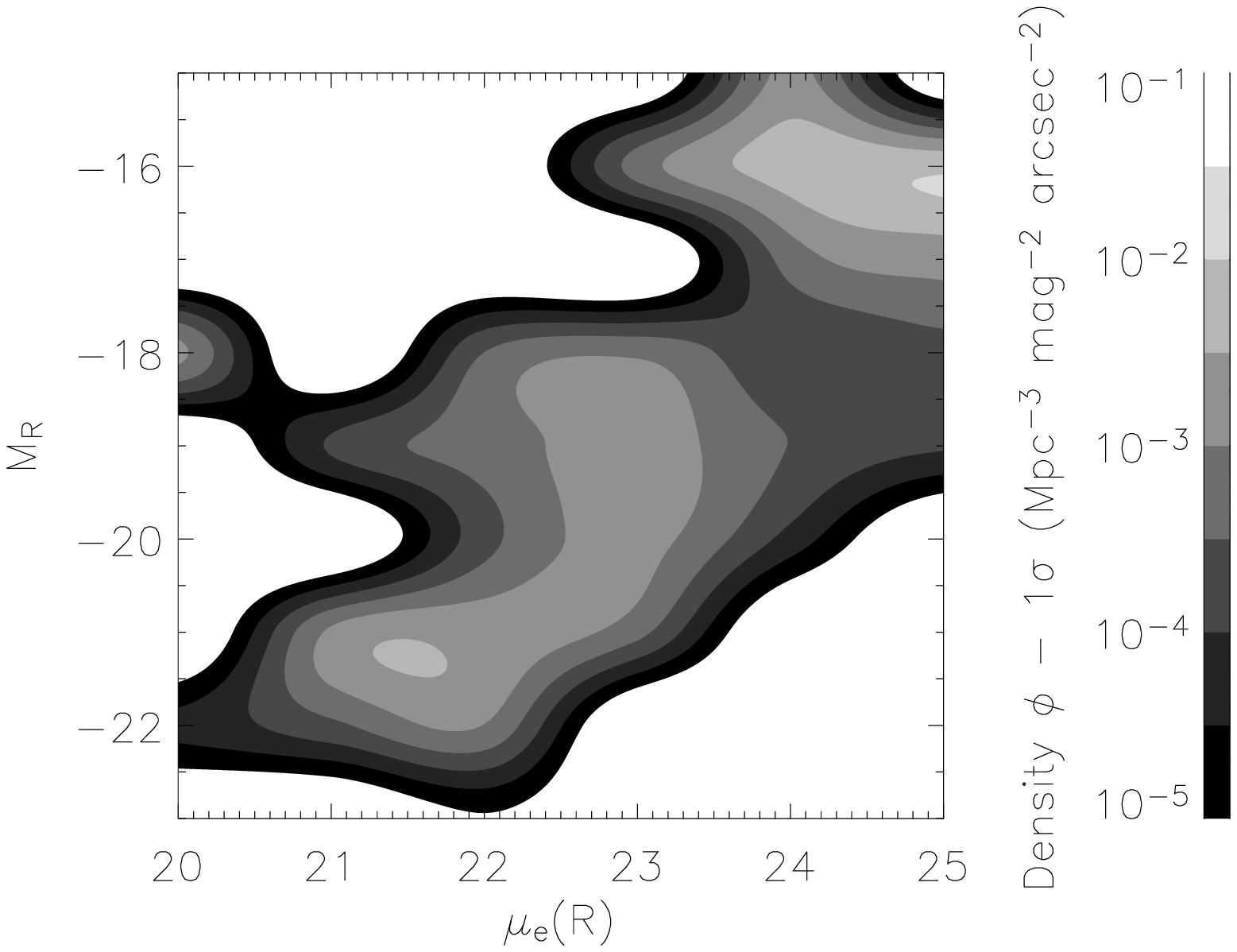}
\hfill
\includegraphics[width=84mm]{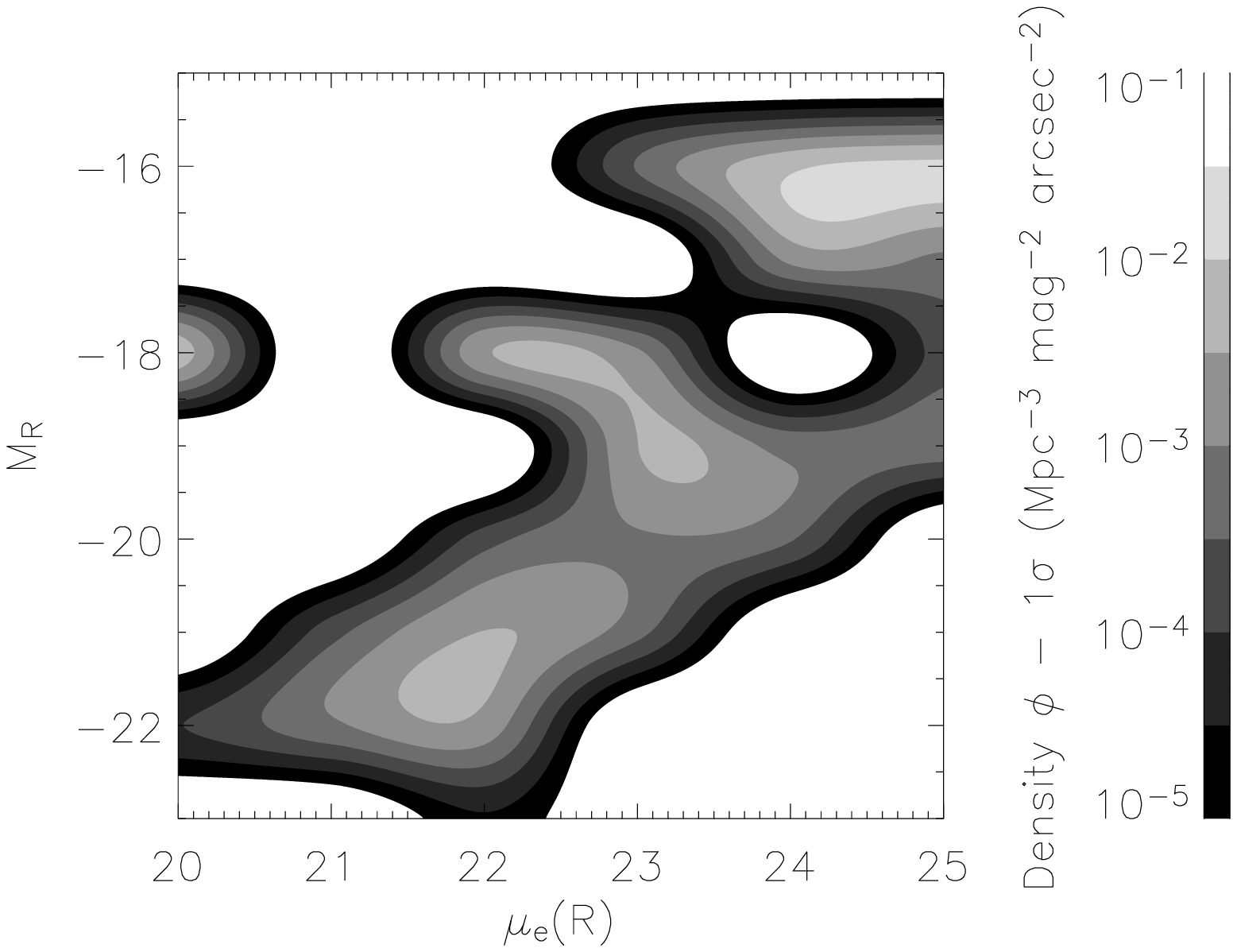}}
\caption{The bivariate brightness distribution (BBD) of H{\sc i}-selected 
galaxies.  The panels to the left show the BBD formed using HIMF-weighting,
as described above, while the panels to the right show the BBD formed using
$1/V_{max}$ weighting.  The central panels give the best-estimate of the
BBD while the top panels show the BBD + 1$\sigma$ in each bin and the 
bottom panels show the BBD - 1$\sigma$ in each bin.}
\label{bbd}
\end{minipage}
\end{figure*}

The corrected BBDs, showing the true space-density of galaxies, are given
in Figure \ref{bbd}.   The HIMF and $1/V_{max}$ corrections give very similar
results.  Over a range of seven magnitudes in luminosity and five in 
surface-brightness, the galaxies appear to lie in an evenly populated strip
four magnitudes wide in luminosity and two in surface-brightness that
stretches from high-luminosity, high-surface-brightness to low-luminosity,
low-surface-brightness.

The HIMF-weighted BBD is obviously dependent on the HIMF used to construct it.
However all of the HIMFs in Table \ref{himfs} give results fairly similar
to those shown here -- if a steeper HIMF is used then the 
density at low-luminosity, low surface-brightness is marginally higher
while if the shallower (and lower-density) HIMF of Zwaan et al. (1997) is 
used then the density decreases all round, with the density of
low-luminosity, low surface-brightness galaxies falling slightly more than
that of high-luminosity, high surface-brightness galaxies.

We define LSB galaxies to be those with effective {\it R}-band 
surface-brightnesses more than 1.5 magnitudes lower than the peak in the 
uncorrected {\it R}-band effective surface-brightness distribution of ESO-LV
galaxies, i.e. $\mu_e (R) > 23.3$ mag arcsec$^{-2}$.  This is slightly dimmer 
than the definitions used by Impey \& Bothun (1997) ($\mu_0 (B) > 23$) and
by McGaugh 
(1996) ($\mu_0 (B) > 22.75$), which are respectively 1.35 and 1.1 magnitudes 
dimmer than the peak in the uncorrected {\it B}-band central surface-brightness
found by Freeman (1970).  Using our definition, we can set limits on the 
population of giant LSB galaxies as follows.

There is a clear deficiency of high-luminosity, LSB `Crouching Giant' 
galaxies (LSB galaxies with $L_R > 10^{10} L_\odot$): of the 47 galaxies with
$L_R > 10^{10} L_\odot$, not one is an LSB galaxy.  By applying the 
binomial theorem, we can calculate that the probability of finding no LSB 
galaxies in a sample of 47 is less than 0.05 if LSB galaxies make up more than 
6 per cent of the population.  We can therefore rule out that LSB galaxies make
up more than 6 per cent of the high-luminosity, gas-rich population with 95
per cent confidence.

Of the sixteen H{\sc i}-massive 
($M_{HI} > 10^{10}M_\odot$) galaxies found in HIDEEP, one
(ESO~383-G059), is an LSB galaxy with $\mu_{e}^R = 23.8$ R$\mu$ -- this galaxy 
has a high H{\sc i} mass despite not having
a high optical ($R$-band) luminosity.  By the same method as above, we
can place a limit on the proportion of H{\sc i} massive galaxies which are
LSB galaxies of 26 per cent  at the 95 per cent confidence level.

There is little evidence that the unpopulated regions at high-luminosity, low
surface-brightness and low-luminosity, high surface-brightness are due to the
H{\sc i} mass limits of the survey (imposed by the down-turn in the H{\sc i}
mass function at high $M_{HI}$ and by small volumes, and the cut at $10^8
M_\odot$, at low $M_{HI}$).  Only one of the galaxies with $M_{HI} > 10^{10}
M_\odot$ is a low surface-brightness galaxy, as previously noted, and the
high-mass galaxies are not spread along the high-luminosity edge of the
populated region as would be expected if this edge were due to an H{\sc i}-mass
selection effect.  Similarly, only one of the galaxies in the lowest mass
range included in the sample ($10^{8.5} M_\odot >M_{HI} > 10^8 M_\odot$) is
a high surface-brightness galaxy (NGC 5253); these galaxies are not spread 
along the low-luminosity edge of the populated region.

The dwarf galaxies that were removed from the sample ($M_{HI} < 10^8 M_\odot$)
fall in the surface-brightness range 22.5 -- 24.5 $R$ mag arcsec$^{-2}$ and
in the $R$ magnitude range -13 -- -18, generally at a lower luminosity than
is `normal' for their surface-brightness.  All but one of these galaxies (all 
in the $1/V_{max}$ sample) lie in the Centaurus A group and are not, therefore,
representative of a variety of environments -- the one dwarf that is not in
the Cen A group (HIDEEP J1337-3118) lies well within the normal range of 
luminosity for its surface-brightness, while the higher-mass galaxy NGC 5253,
which also lies in the Cen A group, is the galaxy most under-luminous for its
surface-brightness in the sample.  Including these dwarf galaxies in the BBD
would not populate in the low-luminosity, high surface-brightness region, but
would extend the BBD to lower luminosities at intermediate and low 
surface-brightnesses.

\begin{figure}
\includegraphics[width=84mm]{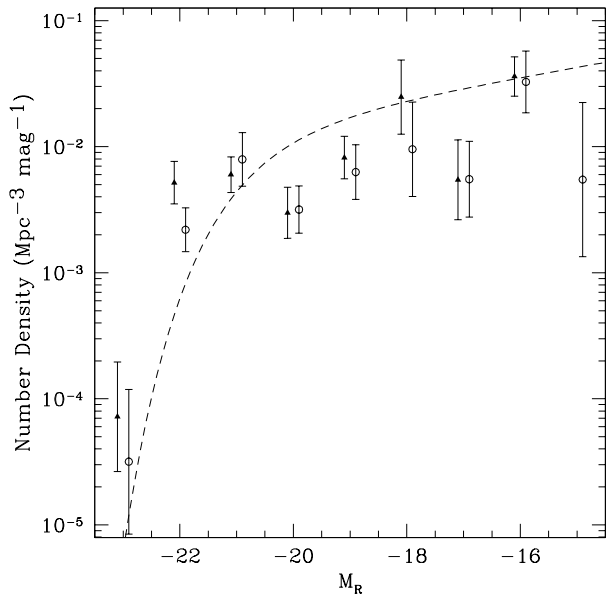}
\caption{Weighted luminosity function of HIDEEP galaxies.  Points
weighted by the HIMF are shown as open circles (offset for clarity by 0.1 mag 
dimmer), points weighted by $1/V_{max}$ are shown as solid triangles (offset 
for clarity by 0.1 mag brighter).  The LF of Blanton et al. (2001, $\alpha = 
-1.20\pm 0.03$) is shown as a dashed line.}
\label{lumfunc}
\end{figure}

By collapsing the BBD along the surface-brightness axis, we can form the
optical Luminosity Function (LF) which is shown in Figure \ref{lumfunc}.
This is
compared to the LF obtained by Blanton et al. (2001) from Sloan Digital Sky 
Survey (SDSS) commissioning data.  It can be seen that both the HIMF and 
$1/V_{max}$ determinations are fairly consistent with the SDSS luminosity 
function, although  $M_\star$ appears to be 
marginally brighter in our data - this could be due to uncertainties in the
conversion between $r^\star$, which Blanton et al. use, and $R$.  The 
$1/V_{max}$ weighting gives a faint-end ($M_R > -21.5$) slope of $\alpha = 
-1.40 \pm 0.14$ and the HIMF-weighting gives $\alpha = -1.27 \pm 0.16$.

\begin{figure}
\includegraphics[width=84mm]{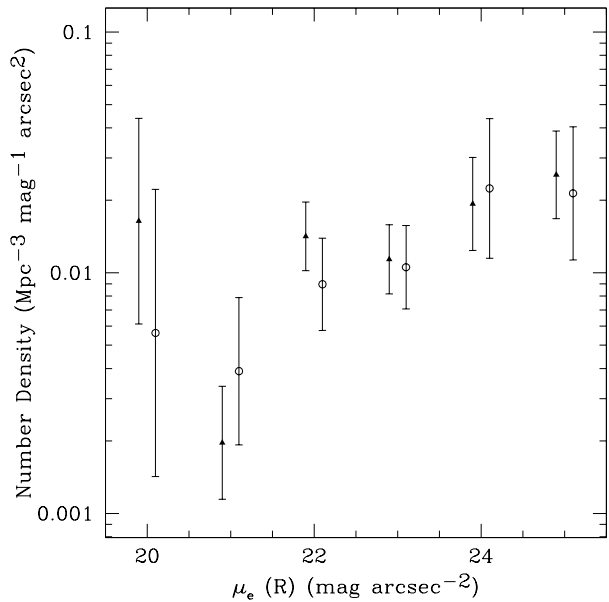}
\caption{Corrected surface-brightness distribution of HIDEEP galaxies. This 
is consistent with a flat or slowly rising surface-brightness distribution at 
lower surface-brightnesses and a down-turn at the bright end.  The 
HIMF-weighted
points are indicated by open circles (offset for clarity by 0.1 mag dimmer)
and the $1/V_{max}$-weighted points by solid triangles (offset for clarity 
by 0.1 mag brighter).}
\label{wtsbd}
\end{figure}

Figure \ref{wtsbd} displays the surface-brightness distribution (SBD), formed 
by collapsing the BBD along the luminosity axis.  This is consistent with
optical determinations such as Davies (1990) and de Jong (1996a).  The
effect of large scale structure on $1/V_{max}$ can be seen in the brightest 
bin.  Both of the galaxies in this bin lie in the close by Cen A group and 
their contribution to the density is therefore over-estimated by this method.
However, in general the $1/V_{max}$ and HIMF-weighted points are in good
agreement, giving confidence that large scale structure is not overly affecting
the $1/V_{max}$ data.

Overall, it can be seen that optical surveys give very similar results to
this survey.  This implies that 21-cm surveys do not uncover any `hidden' 
population of extremely low surface-brightness galaxies that is missed by
 optical 
surveys.  If such a population does exist in significant numbers, it must be 
composed primarily of galaxies with neutral gas masses lower than 
$10^8 M_\odot$.

This is not to say, however, that LSB galaxies do not make up a significant
population.  The SBD derived from weighting by the BGC HIMF implies that
LSB galaxies make up 62 $\pm$ 37 per cent of the total population of galaxies
with $M_{HI} > 10^8 M_\odot$, or 51 $\pm$ 20 per cent for $1/V_{max}$ 
weighting.  Even with the large errors on these estimates, it is clear that 
a large number of 
galaxies fall into our definition of low surface-brightness.  In the next
section we investigate what contribution these LSB galaxies make to the 
Universe.

\section{The cosmological significance of low surface-brightness galaxies}
\label{cosmo-sec}

As LSB galaxies have been proposed as repositories for some of the missing
baryons (e.g. Impey \& Bothun 1997) and may also contain large quantities of 
dark matter, it is important to make an estimate of how much they actually
contribute to the Universe.  It is possible to do this from our data, subject 
to the provisos 
that H{\sc i}-poor galaxies would not be found at 21-cm (e.g. elliptical
galaxies) and that we do not detect sufficient numbers of dwarf galaxies 
($M_{HI} < 10^8 M_\odot$) to say anything about their contribution.  We compare
the contribution of the LSB galaxies in our sample to the total contribution
of H{\sc i}-rich galaxies
using the weightings described in Section \ref{volume-sec} to correct the 
numbers in each surface-brightness bin. 

\begin{figure}
\includegraphics[width=84mm]{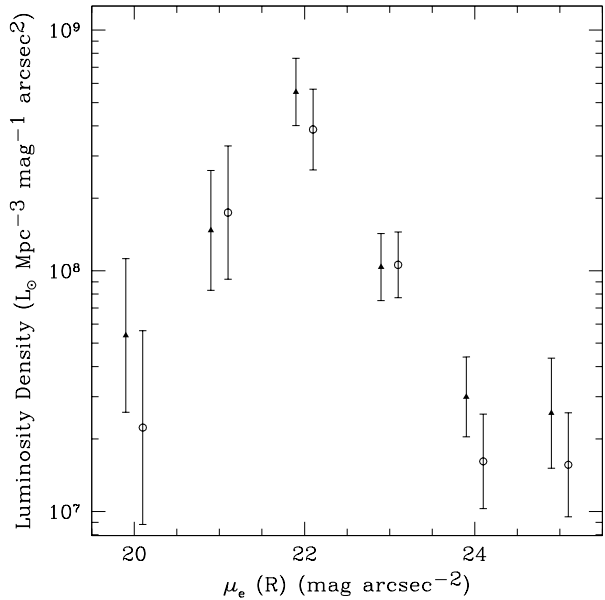}
\caption{Luminosity density -- surface-brightness distribution for 
HIDEEP galaxies.  The luminosity density can be seen to fall 
sharply either side of a peak near the Freeman-law value.  The HIMF-weighted
points are indicated by open circles (offset for clarity by 0.1 mag dimmer)
and the $1/V_{max}$-weighted points by solid triangles (offset for clarity by 
0.1 mag brighter).}
\label{lumdensity}
\end{figure}

To compare the contribution to the luminosity-density made by galaxies of 
different surface-brightnesses we need to weight the surface-brightness 
distribution in Figure \ref{wtsbd} with the luminosities of the galaxies.
When this is done we obtain Figure \ref{lumdensity} 
which shows that the luminosity density is sharply peaked near the Freeman-law
value.  Gas-rich LSB galaxies do not appear to emit much light: when analysed
with the HIMF-weighting, they contribute $6.7\pm 2.8$ per cent of the total 
luminosity-density of all gas-rich galaxies, or $6.5 \pm 2.4$ per cent 
according to the $1/V_{max}$ analysis.  This is very similar to the $7.3 \pm
3.6$ per cent contribution for LSB galaxies found by Driver (1999) for 
local ($0.3 < z < 0.5$) galaxies in the Hubble Deep Field.  The main source
of errors in our determination is Poisson noise in the surface-brightness
bins but there is also a contribution due to the uncertainty on the luminosity
of the sources and, for the H{\sc i} mass function weighting, due to the 
width of the mass bins and the number of galaxies in each bin.

\begin{figure}
\includegraphics[width=84mm]{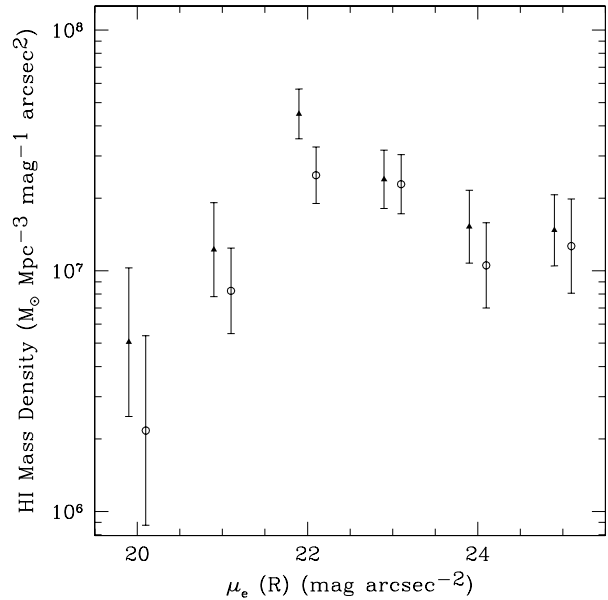}
\caption[Neutral hydrogen density -- surface-brightness distribution for
HIDEEP galaxies]{Neutral hydrogen density -- surface-brightness 
distribution for HIDEEP galaxies.  It can be seen that this is consistent
with a slowly falling distribution towards lower surface-brightnesses and a 
much steeper fall-off
towards higher surface-brightnesses, similar in shape to the surface-brightness
distribution.  The HIMF-weighted
points are indicated by open circles (offset for clarity by 0.1 mag dimmer)
and the $1/V_{max}$-weighted points by solid triangles (offset for clarity 
by 0.1 mag brighter).}
\label{hidensity}
\end{figure}

Similarly the contribution of LSB galaxies to the H{\sc i} content as a whole 
can be calculated and is shown in Figure \ref{hidensity}.  This amounts to 
$32\pm 11$ per cent of all H{\sc i} with HIMF-weighting, or $27 \pm 7$ per cent
with $1/V_{max}$ weighting.

This is inconsistent, at the 1$\sigma$ level, with the determination of Zwaan
et al. (2003) that LSB galaxies contribute only 15 per cent to the total
H{\sc i} mass density.  That determination should, however, be treated as a
lower limit.  The sample of Zwaan et al., unlike our sample, did not have
complete optical data -- surface-brightnesses were only available for 600
out of the 1000 galaxies.  When corrections for this were made, Zwaan et
al. assumed that the incompleteness in the optical data was unrelated to
surface-brightness.  It is far more likely that the LSB galaxies' data were 
more incomplete than the HSB galaxies', which could raise the contribution of
LSB galaxies considerably.

The baryonic content of the galaxies was calculated following the method of
McGaugh et al. (2000), adding the mass of the stars and the gas together 
to get a total baryonic mass for the galaxy:

\begin{equation}
M_{bary} = 1.4 M_{HI} + \Upsilon^X_\star L_X
\end{equation}

\noindent where $\Upsilon^X_\star$ is the stellar mass to light ratio in
the band used, and $L_X$ is the luminosity in that band.  For our R-band
data,  $\Upsilon^R_\star$  has been estimated, as in McGaugh et al., using 
the model of de Jong (1996b) for a 12 Gyr old, solar metallicity stellar 
population with a constant star formation rate and a Salpeter initial
mass function, corrected to {\it R}-band using the average colours in 
that paper (also used by McGaugh et al. for their correction to 
{\it H}-band).  This gives a value of $\Upsilon^R_\star \approx 1.4$, 
which we have used in our calculations.

The relative contribution of LSB galaxies to the total baryon density 
is then calculated to be $9.3\pm 3.6$ per cent using HIMF-weighting or
$8.7 \pm 2.9$ per cent using $1/V_{max}$.  This is only marginally higher
than the contribution to the luminosity, reflecting that the contribution
to H{\sc i} mass density is basically flat and so only slightly affects
the shape of the density distribution when the two are added together.  LSB
galaxies do have more of their baryons in the form of gas, as shown in the
relationship between $M_{HI}/L$ and surface-brightness, but this is outweighed
by their much lower luminosities.

\begin{figure}
\includegraphics[width=84mm]{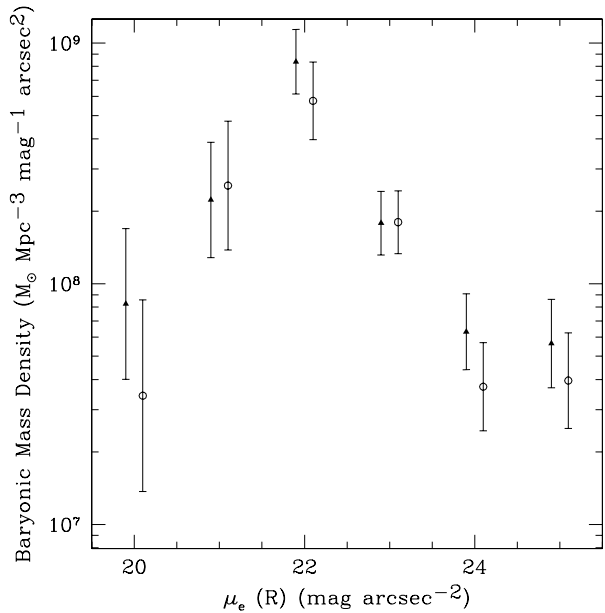}
\caption[Baryon density -- surface-brightness distribution for HIDEEP
galaxies]{Baryon density -- surface-brightness distribution for HIDEEP
galaxies.  It can be seen that the greatest contribution to the baryon
density is made by Freeman-law galaxies around $\mu_{e}^\star$, with
the density falling of towards lower and higher surface-brightnesses.  The 
HIMF-weighted points are indicated by open circles (offset for clarity by 0.1 
mag dimmer) and the $1/V_{max}$-weighted points by solid triangles (offset 
for clarity by 0.1 mag brighter).}
\label{barydensity}
\end{figure}

\begin{figure}
\includegraphics[width=84mm]{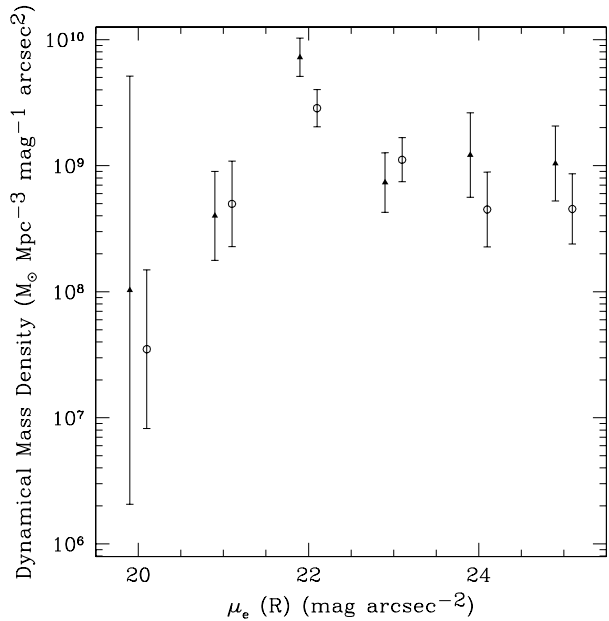}
\caption{Mass density -- surface-brightness distribution for HIDEEP 
galaxies.  The greatest contribution is made by Freeman-law
galaxies $\mu_{e}^\star$, however the distribution seems fairly flat towards
lower surface-brightnesses while it falls off towards higher 
surface-brightnesses.  The HIMF-weighted
points are indicated by open circles (offset for clarity by 0.1 mag dimmer) 
and the $1/V_{max}$-weighted points by solid triangles (offset for clarity by 
0.1 mag brighter).}
\label{massdensity}
\end{figure}

If, in the usual way, we estimate the dynamical masses of HIDEEP galaxies using
\begin{equation}
M_{dyn} = \frac{R_{HI}\times(\Delta V_0)^2}{\hbox{G}}
\end{equation}
\noindent where we follow Paper 1 in assuming 
$r_{HI}\simeq 5 r_{e}$
and $\Delta V_0$ is the inclination-corrected velocity width ($\Delta V_0 =
\Delta V/\sin i$) then we arrive at the relative contribution
of galaxies of various surface-brightnesses shown in Figure \ref{massdensity}.
For this calculation, only those galaxies with reliable inclinations in the
range $45^\circ < i < 80^\circ$ were used.  This leaves 57 galaxies in the
HIMF-weighted sample and 38 in the $1/V_{max}$-weighted sample.
The share due to LSB galaxies is, although uncertain, relatively high at 
$22\pm 10$ per cent for HIMF-weighting or $21\pm 12$ per cent for 
$1/V_{max}$-weighting.  This is in keeping with other studies that have shown 
that LSB galaxies contain proportionally more dark matter than normal galaxies.

\begin{figure}
\includegraphics[width=84mm]{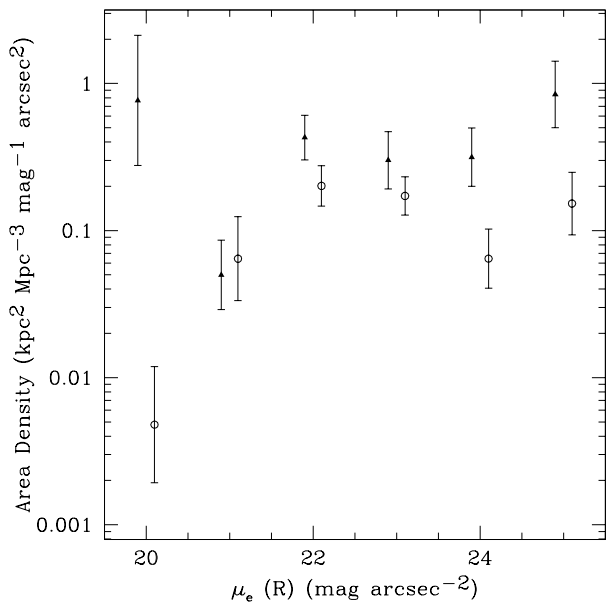}
\caption{Cross-sectional area density against surface-brightness
for HIDEEP galaxies.  This is fairly flat, with a fall off towards higher
surface-brightnesses (the brightest point of the $1/V_{max}$ distribution is,
as noted earlier, affected by large scale structure).  This implies that 
LSB galaxies will make a significant contribution to QSO absorption lines in
so far as these are caused by galaxies.  The HIMF-weighted
points are indicated by open circles (offset for clarity by 0.1 mag dimmer)
and the $1/V_{max}$-weighted points by solid triangles (offset for clarity by 
0.1 mag brighter).}
\label{xsecdensity}
\end{figure}

Since the cross-sections of large, luminous  galaxies can by no means explain 
the prevalence of quasar absorption line systems (QSOALS) it has been suggested
that LSB galaxies might be responsible (e.g. Phillipps et al. 1993, Linder 
1998; 2000) though this has proved controversial (e.g. Chen et al. 1998).  
If we assume that the absorption cross-sections of our galaxies are 
proportional to their effective areas (e.g. the area enclosed by the
effective radius) then we can make an estimate of the contribution of LSB
galaxies to this cross-section.  Figure \ref{xsecdensity} shows the 
cross-sectional area distribution of our galaxies with surface-brightness
(corrected to an inclination of 60$^\circ$), formed in the same way as the 
above plots This gives the contribution of LSB galaxies to the cross-section to
be 39 $\pm$ 15 per cent using the HIMF-weighting and 42 $\pm$ 22 per cent using
$1/V_{max}$-weighting.

Their high cross-section suggests that LSB galaxies are likely to form
a significant fraction of the absorbers where galaxies themselves are 
responsible for QSOALS, as they might be in the case of damped Lyman-$\alpha$
systems.  This is born out by recent observations where LSB galaxies, rather
than HSB galaxies with immense halos, appear to be the more likely absorbers
(Turnshek et al. 2001; Bowen, Tripp \& Jenkins 2001).
 
\begin{table*}
\begin{minipage}{146mm}
\caption{Summary of the measured contribution of HSB and LSB galaxies to the
density of various quantities in gas-rich galaxies.  These are the
actual measured densities in the HIDEEP survey, for the SBDs constructed
using weighting by the HIMF of Zwaan et al. (2003) and using $1/V_{max}$ 
weighting.  Possibly systematic errors due to the selection of the HIMF, 
calculated by comparing the results of the Zwaan et al. (2003) HIMF with other
published HIMFs as described earlier, are given as a second error to the
percentage LSB contribution.}
\label{importantLSB2}
\begin{tabular}{lllrrr}
Quantity&Weighting&N$_{gal}$&HSB contribution&LSB contribution&Percentage LSB 
contribution\\
\hline
Number density&HIMF&101&1.8 $\pm$ 0.7&3.0 $\pm$ 1.5&62 $\pm$ 37 $^{+5}_{-3}$\\
(10$^{-2}$ galaxies Mpc$^{-3}$)&$1/V_{max}$&62&2.9 $\pm$ 1.1&3.0 $\pm$ 0.9&
51 $\pm$ 20\\
Neutral Hydrogen density&HIMF&101&3.7 $\pm$ 0.7&1.8 $\pm$ 0.5&32 $\pm$ 11
$^{+10}_{-3}$\\
($10^7M_\odot$ Mpc$^{-3}$)&$1/V_{max}$&62&5.7 $\pm$ 1.2&2.1 $\pm$ 0.5&
27 $\pm$ 7\\
Luminosity density&HIMF&101&45 $\pm$ 13&3.2 $\pm$ 1.0&7 $\pm$ 3 $^{+1}_{-1}$\\
 ($10^7L_\odot$ Mpc$^{-3}$)&$1/V_{max}$&62&57 $\pm$ 13&3.9 $\pm$ 1.2&
7 $\pm$ 2\\
Baryon density&HIMF&101&68 $\pm$ 17&6.9 $\pm$ 2.1&9 $\pm$ 4 $^{+2}_{-1}$\\
($10^7M_\odot$ Mpc$^{-3}$)&$1/V_{max}$&62&88 $\pm$ 20&8 $\pm$ 2&
9 $\pm$ 3\\
Mass density&HIMF&57&280 $\pm$ 80&77 $\pm$ 31&22 $\pm$ 10 $^{+4}_{-2}$\\
($10^7M_\odot$ Mpc$^{-3}$)&$1/V_{max}$&38&570 $\pm$ 170&150 $\pm$ 78&
21 $\pm$ 12\\
Area density&HIMF&101&5.4 $\pm$ 1.2&3.4 $\pm$ 1.2&39 $\pm$ 15 $^{+8}_{-3}$\\
($10^{-1}$ kpc$^2$ Mpc$^{-3}$)&$1/V_{max}$&62&7.4 $\pm$ 1.7&5.4 $\pm$ 2.5&
42 $\pm$ 22\\
\end{tabular}
\end{minipage}
\end{table*}

Table \ref{importantLSB2} summarises the findings of this section.  It can be 
seen that LSB galaxies make up over half of all gas-rich galaxies, yet have 
less than
10 per cent of the luminosity density.  The relationship between dynamical 
$M/L$ and surface-brightness means that luminosity is a biased indicator of 
the cosmological significance of LSB galaxies.  Similarly, the higher H{\sc i} 
mass-to-light ratios of LSB galaxies means that they have more gas than would 
be indicated by their light on a straight extrapolation of 
$M_{HI}/L$ from Freeman-law galaxies.  The relatively larger sizes of
LSB galaxies means that their contribution to cross-sectional area,
around 40 per cent, is much higher than would be expected from their luminosity
or even from their H{\sc i} mass.

The totals given here are only for galaxies with $M_{HI} > 10^8 M_\odot$.
These are almost entirely spiral galaxies.  Dwarf galaxies, even gas-rich
ones, tend to have lower  gas masses than this, while elliptical galaxies
are too gas-poor to be detected.  Most dwarf galaxies have low 
surface-brightnesses, therefore if these were included it is likely that
the total contributions from LSB galaxies would rise.  These numbers should
therefore be seen as lower estimates for the total contribution of LSB
galaxies.

\section{Conclusions}
\label{conc-sec}

This survey does not find any LSB `Crouching Giant' galaxies, such as Malin 1.
From this, we can rule out LSB galaxies making up more that 6 per cent of the
population of luminous ($L_R > 10^{10} L_\odot$) galaxies.  Indeed, there 
does seem to be a minimum surface-brightness at every magnitude level given
approximately by $\mu_{e,min} = 45 + M_R$ (Figure \ref{bbd}).  To higher
surface-brightnesses, galaxies seem to populate fairly evenly a band 
approximately 4 magnitudes wide in luminosity and 2 in surface-brightness.  
Beyond this, high surface-brightness, low-luminosity galaxies also appear to 
be rare in the gas-rich population.  The populated band broadens slightly 
towards lower luminosities, giving it an approximate slope of $\Sigma \propto 
L^{0.7}$ (where $\Sigma$ is surface brightness in luminosity per unit area and 
$L$ is the total luminosity).  From this, it can be calculated that the radius,
$R$, is related to the surface-brightness as $R \propto \Sigma^{0.2}$, i.e. the
radius only changes very slightly with surface-brightness.

Once volumetric corrections
are made, the number of galaxies per unit volume is flat as we go to lower 
surface-brightnesses.  Even within the surface-brightness limit reached by
this survey, LSB galaxies contribute over half of the number density of
galaxies.   Furthermore, we find that LSB galaxies contribute approximately
30 per cent of the neutral-hydrogen density, twice the lower limit set by Zwaan
et al. (2003).  LSB galaxies may also contribute around 20 per cent of the 
total mass density (again with large errors), but only 7 per cent of the 
luminosity-density.

Overall, H{\sc i}-rich LSB galaxies do not appear to contribute much to the 
Universe.  However, the cross-section to QSOALS made by LSB galaxies, which is 
around 40 per cent of the total cross-section, is disproportionate to their 
luminosity or baryonic content. This implies that LSB galaxies could contribute
substantially to those QSOALS where galaxies are the absorbers.

Future work in progress will see an order of magnitude larger sample assembled 
from the overlap between HIPASS and the SDSS.  This will give not only an 
increase
in the significance of the results but will also give 5-colour optical data,
thus greatly extending our knowledge of H{\sc i}-selected galaxies.  A larger
sample of H{\sc i}-selected samples from the SDSS may be assembled in the 
medium term by the proposed ALFALFA drift-scan survey with the Arecibo L-band 
Feed Array or in the longer term by surveys carried out with the Square
Kilometre Array.  However, as neither HIPASS nor ALFALFA reach column-density
levels as low as HIDEEP it is unlikely that they will turn up any
significant population of extremely LSB galaxies that we have missed.

This is not to say that extremely LSB galaxies do not exist, just that if they
do make up a significant population then the amount of H{\sc i} they contain
must be small.  In order to search for these galaxies, techniques other than 
H{\sc i}
surveys will be necessary.  The optical channel available to us on the ground
is fundamentally unsuited to looking for LSB galaxies, but a quite modest
switch in wavelengths to either side of the solar peak, which is possible from
space, could yield dramatic results.  O'Connell (1987) has discussed the idea
of using UV wavelengths, while in the $H$-band at 1.6 $\mu$m the contrast
between the Sun's scattered zodiacal spectrum and the light from red stars in
LSB galaxies could be two orders of magnitude higher than the contrast from
the ground.  Wide-field near-IR cameras, such as the WFC-3 instrument currently
awaiting shipment to the HST, will be needed to exploit this window which is,
in principle, capable of revealing galaxies 7 magnitudes lower in 
surface-brightness than we can detect from the ground.

The authors would like to thank Lister Staveley-Smith, Tony Fairall, David
Barnes, Jon Davies and Suzanne Linder for useful discussions.  RFM, WJGdeB
and PJB acknowledge the support of PPARC.  BGK acknowledges the support of the
Australian Research Council.  The authors would also like to thank the 
staff of the CSIRO Parkes and Narrabri observatories, of the NRAO Array 
Operations Center and of the ANU Siding Spring Observatory for their help 
with observations.  We
acknowledge PPARC grant GR/K/28237 to MJD towards the construction of the
multibeam system and PPARC grants PPA/G/S/1998/00543, PPA/G/S/1998/00620,
PPA/G/S/2002/00053 and PPA/G/S/2003/00085 to MJD towards its operation.  We
would particularly like to thank the anonymous referee for an extremely 
helpful report.

This research has made use of the NASA/IPAC Extragalactic Database (NED) which
is operated by the Jet Propulsion Laboratory, Caltech, under agreement with 
the National Aeronautics and Space Administration.  This research has also 
made use of the Digitised Sky Survey, produced at the Space Telescope Science
Institute under US Government Grant NAG W-2166 and of NASA's Astrophysics 
Data System Bibliographic Services

\end{document}